\documentstyle[aps]{revtex}
\begin{document}
\draft
\preprint{}
\title{Evidence for ideal insulating/conducting state in a 1D integrable
system}
\author{X. Zotos}
\address{
Institut Romand de Recherche Num\'erique en Physique des
Mat\'eriaux (IRRMA), \\
PHB-Ecublens, CH-1015 Lausanne, Switzerland}
\author{P. Prelov\v sek}
\address{J. Stefan Institute, University of Ljubljana, \\
61111 Ljubljana, Slovenia}
\date{Received\ \ \ \ \ \ \ \ \ \ \ }
\bigskip\bigskip
\maketitle
\begin{abstract}
Using numerical diagonalization techniques we analyze the finite
temperature/frequency conductance of a one dimensional model of interacting
spinless fermions. Depending on the interaction,
the observed {\it finite temperature} charge stiffness
and low frequency conductance indicate a fundamental
difference between integrable and non-integrable
cases.  The integrable systems behave as ideal conductors in the
metallic regime and as ideal insulators in the insulating
one. The non-integrable systems are, as expected, generic conductors in
the metallic regime and activated ones in the insulating regime.
\end{abstract}


%\narrowtext
In classical integrable systems the existence of a macroscopic number
of conservation laws has a profound consequence, dissipationless
transport\cite{sol}.  The analogous effect
in nontrivial {\it quantum} integrable systems has been studied only
recently in a prototype model of dissipation\cite{czp}. In this model,
describing a particle interacting
with a fermionic bath in one dimension, we found that
the tagged particle shows ideal mobility even at finite temperatures, $T>0$,
when the system is integrable.
It is desirable to extend these ideas to homogeneous
many-body models, which are at present of particular interest in
connection with strongly correlated electrons. In this field, most analytical
findings are on integrable 1D models as the Hubbard\cite{lw} or the
spinless fermions with nearest neighbor interaction model\cite{emery}.

Progress in the study of dynamical response at $T>0$ is
hindered by the lack of reliable methods. The only attempts,
which however might obscure the role of integrability,
start from a Luttinger liquid effective Hamiltonian
description\cite{gm}.

In this work, based on a recent reformulation of the finite temperature
charge stiffness\cite{czp} and numerical methods for calculating
$T>0$ dynamical conductivities $\sigma(\omega)$ \cite{pp} on finite
size systems, we present evidence of the importance of integrability
for transport properties.

{}From linear response theory, the real part of the conductance
$\sigma(\omega)$ at frequency $\omega$ is given by:
\begin{equation}
\sigma(\omega)= 2\pi D\delta(\omega)+\sigma_{reg}(\omega), \label{eq}
\end{equation}

\begin{equation}
\sigma_{reg}(\omega)=
\frac{1-e^{-\beta \omega}}{ \omega} \frac{\pi}{L}
\sum_{n,m\neq n} p_n  \mid \langle n \mid \hat j \mid m \rangle \mid ^2
 \delta(\omega- \epsilon_m + \epsilon_n), \label{eq1}
\end{equation}
where $\mid n\rangle, \epsilon_n$ denote the eigenstates and
eigenvalues of the Hamiltonian, $p_n$ the corresponding Boltzmann
weights, $\hat j$ the current operator and $\beta=1/k_B T$. We will
consider 1D tight-binding models of $L$ sites ($k_B=\hbar=e=1$).

$\sigma(\omega)$ satisfies the optical sum rule\cite{mald}:
\begin{equation}
\int_{-\infty}^{\infty} \sigma(\omega) d\omega =\frac{\pi}{L}
\langle -\hat T \rangle, \label{eq3}
\end{equation}
where $\langle \hat T\rangle$ denotes the thermal expectation value of the
kinetic energy. The sum rule (\ref{eq3}) together with
Eqs.(\ref{eq},\ref{eq1}) can be used for the evaluation of the
stiffness $D(T)$ \cite{Kohn,czp}.  It will be however more convenient to
discuss the behavior of $D$ at finite temperatures, with a generalization
of the original Kohn's
approach \cite{Kohn} for zero temperature, by relating $D(T)$ to the thermal
average of curvatures of energy levels subject to a fictitious flux $\phi$
\cite{czp} :
\begin{equation}
D = \frac{1}{L} \sum_n p_n D_n= \frac{1}{L} \sum_n p_n
\frac{1}{2} \frac {\partial ^2 \epsilon_n (\phi) }{\partial \phi^2}|_{\phi
\rightarrow 0} .
\label{eq4}
\end{equation}

At zero temperature $D(T=0)=D_0$ has been introduced \cite{Kohn} to
distinguish an
ideal conductor with $D_0>0$ from an insulator with $D_0=0$.  Our aim here is
to analyze the transport behavior at finite temperatures. For orientation,
at $T>0$,  a conductor can
develop either to a {\it normal conductor} (resistor) with $D(T)=0$
but $\sigma_0=\sigma(\omega\rightarrow 0)>0$, or remain an
{\it ideal conductor}
characterized by $D(T)> 0$.  An insulator might develop to a
normal conductor (conducting by thermally activated transport)
with $D(T)=0, \sigma_0(T)>0$, remain an {\it ideal insulator} with
$D(T)=0, \sigma_0(T)=0$ or even become an ideal conductor with $D(T)>0$.

Below we present results for $\sigma(\omega)$ for the prototype 1D
tight-binding model of interacting spinless fermions with nearest and
next-nearest neighbor interaction.
For systems with Hilbert space dimension less than typically 1000
states (after implementation of translational symmetry) we calculate
$\sigma(\omega)$ directly from Eq.(\ref{eq1}) by finding all
eigenstates and evaluating current matrix elements; for systems with
larger basis dimensions we use a $T>0$ Lanczos-based numerical
technique \cite{pp}.

As we are interested mostly in
differences in the {\it qualitative behavior} of integrable
vs. non-integrable systems we can restrict our study to
high temperatures thus minimizing spurious effects due to the sparse
low energy level spectrum in finite size systems.
It corresponds, in normal conductors, studying systems
with mean free paths shorter than the lattice size.

Further, we present the integrated and normalized intensity
\begin{equation}
I(\omega)=D^* +
\frac{2L}{\langle -\hat T \rangle}
\int_{0^+}^{\omega} d\omega' \sigma(\omega') ; ~~
D^*=\frac{2LD}{\langle -\hat T \rangle}
\end{equation}
as it contains the relevant information in the conductance and avoids
the smoothing procedure of the discrete $\sigma(\omega)$ spectra of finite
size systems.

The Hamiltonian we study is given by:
\begin{equation}
\hat H=-t\sum_{i=1}^L (e^{i\phi}c^{\dagger}_{i+1} c_{i} + H.c.)+
V\sum_{i=1}^L n_i n_{i+1} + W \sum_{i=1}^L n_i n_{i+2} \label{eq5}
\end{equation}
where $c_i(c^{\dagger}_i)$ are annihilation (creation) operators of a
spinless fermion at site $i$, $n_i=c_i^{\dagger}c_i$.
This Hamiltonian is integrable using the Bethe ansatz method for $W=0$
\cite{emery} and non-integrable for $W\ne 0$. For
$W=0$ and $V< 2t$ the ground state is metallic, while for $V> 2t$ a
charge gap opens and the system is an insulator.

We study numerically various size systems with periodic boundary conditions
and $M=L/2$ fermions (half-filling).  The results for
$L=8,12,16$ are obtained by the complete diagonalization of the
Hamiltonian, while for $L=20,24$ the Lanczos method is employed.
It should be mentioned that in the latter cases results,
e.g. for $D^*$, are subject to small statistical error due to finite
random sampling\cite{pp}.

{\it Metallic state:} In Fig.~1 we show the finite temperature
conductance for an integrable
case. To study the finite size dependence of the charge stiffness,
we plot in the inset $D^*$ as a function of $1/L$;
the dashed lines indicate a $3^{rd}$ order polynomial
extrapolation based on the $L=8,12,16$ site systems, suggested by the
very good agreement obtained with the $T=0$ analytical result (square
at $1/L=0$)\cite{ss}.  We find that for
$L\rightarrow \infty$ the extrapolated $D^* \neq 0$. At the same time
$\sigma_0=\sigma(\omega\rightarrow 0)\rightarrow 0$
as $I(\omega)$ seems to
approach $\omega=0$ with zero slope ($\sigma(\omega)$ is the derivative of
$I(\omega)$). This behavior is reminiscent of a
pseudogap. These two results indicate that the integrable
system behaves as an {\it ideal conductor} at $T>0$. Moreover we find that
the normalized $D^*$ approaches a nontrivial finite value
$D^*_{\infty}$ in the limit $T\rightarrow \infty$,
depending on $V/t$ and filling,
as both $D$ and $\langle -\hat T\rangle$ are proportional to $\beta$
in this limit.

In Fig.~2 we show $I(\omega)$ and $D^*$ for a non-integrable case.
Here, as expected for a generic metallic
conductor (resistor), we find that $D^*$ scales to zero, probably
exponentially with system size, and
$\sigma_0 > 0$ as $I(\omega)$ approaches
$\omega=0$ with a finite slope.  These two results imply that the
non-integrable system behaves as a normal conductor at $T>0$.

To further point out the particularity of integrable systems, we
investigate the behavior of the conductance
on approaching the integrable point $W=0$.
In Fig.~3 we present $I(\omega)$ scanning the parameter $W$.
We clearly recognize a continuous
transfer of the $\delta-$function weight $I(\omega=0)=D^*$ at $W=0$ to
low frequencies, both for $W> 0$ and $W<0$. From a calculation of
the second frequency moment of the conductance at infinite temperature,
we estimate the frequency range of $\sigma(\omega)>0$ proportional to
$(V-W)^2+W^2/2t^2$. Due to remaining finite size effects we are not
attempting yet to make more quantitative statements about the critical
behavior of the low frequency conductance.

These numerical results on the finite temperature charge stiffness, although
not conclusive, strongly suggest a qualitative difference between integrable
and nonintegrable systems. We can argue about this difference by considering
the expression for $D$ as a thermal average of the curvatures of levels
subject to a fictitious flux. The integrable systems, as they are
characterized by absence of repulsion between levels, they allow larger
fluctuations in the level response to a flux and thus plausibly a finite
charge stiffness.

On the other hand in the nonintegrable systems, because of level repulsion,
the motion of levels with flux is constrained within a characteristic level
spacing that is proportional to $e^{-\alpha L}$, the density of states,
and thus to a vanishing charge stiffness with increasing system size.

We have also verified that in our integrable system the absence of level
repulsion leads to Poisson statistics of the level spacings while
in the integrable one the level repulsion leads to GOE statistics\cite{bell}.

As for $\sigma_0$, it is more difficult to
ascertain its behavior in the infinite size limit from
numerical results in finite size systems.
For the nonintegrable systems we find, as expected, a finite value
for $\sigma_0$. For the finite size integrable systems that we can
study, although $I(\omega)$ seems to
approach $\omega=0$ with a zero slope, we cannot really exclude a finite
slope for $L\rightarrow\infty$. However, from the physical point of view,
even in this case one can expect ideal conductance provided
the charge stiffness remains finite.  It is indicative of a
free accelerating system similar in the spirit of a two fluid model.

{\it Insulating state:} In Fig.~4 we show $I(\omega)$ for the
integrable case $V=4t, W=0$.
At this value of $V>2t$ the ground state is insulating
characterized by $D_0=0$ and a charge gap $\Delta_0\simeq t$\cite{cg}.
We find that, at finite temperatures, $D^*(T>0)=I(\omega =0)$ seems also
to {\it decrease} exponentially with the system size
scaling to zero for $L\rightarrow \infty$. This precludes a possibility
for ideal conductance at $T>0$. Furthermore $I(\omega)$ seems to approach
$\omega=0$ with zero slope, showing a depletion of weight within
a low frequency region of order $\omega < t$.  These are
characteristics of an {\it ideal insulator}, not conducting even at
high temperatures $T \gg \Delta_0$.

In contrast, as shown in the inset of Fig.~4 (for $L=16$),
non-integrable systems of roughly the same charge gap
exhibit a qualitatively different behavior. $I(\omega)$ approaches
$\omega=0$ with finite (although small) slope, consistent with a small
static conductance $\sigma_0>0$. As expected, conductance here is of a
thermally activated character, since appreciable $\sigma_0>0$ appears
only at elevated $T\gg t$.

To discuss the behavior of the conductance in the insulating regime, it is
a good starting point to think about the large $V/t$ limit.
In this limit the energy spectrum splits in "Hubbard" bands with a fixed
number $N_s$ of soliton-antisoliton ($s\bar s$) pairs,
solitons corresponding to nearest neighbor occupied sites while
antisolitons to nearest neighbor empty sites. In the $V=\infty$ limit
solitons and antisolitons are impenetrable. Crossing can only occur by
annihilation and creation of a pair which corresponds to mixing with other
"Hubbard" bands and it has an amplitude of order $t^2/V$.
The states are grouped in characteristic sequences of solitons and antisolitons
and the energies of the eigenstates are the same as those of a gas of 2$N_s$
impenetrable particles.

To analyze the flux $\phi$ dependence of the energy we note
that the phase associated with the hopping of a
soliton in a given direction is opposite to the phase picked by an antisoliton;
so solitons and antisolitons carry opposite charge.
It follows that, in this $V=\infty$ limit, the flux $\phi$ dependence
of the hopping matrix elements can be removed by a redefinition of the
phase of the states.
An nontrivial $\phi$ dependence would have appeared if by successively
applying the Hamiltonian on a given $s\bar s$ state we could bring it to
an equivalent one but with an accumulation of a nonzero phase factor. This
process, which corresponds to a uniform translation of fermions, is not
possible provided we do not allow for soliton-antisoliton crossing.

Therefore, as we have also numerically verified,
the width of the "Hubbard" band is of the order of $N_s t$
but the energy levels are independent of the
flux $\phi$ and $D$ is strictly zero. At the same time it is
reasonable to argue that a static electric field acting on
impenetrable particles of opposite charge cannot produce a uniform current
and the static conductance $\sigma_0$ should also be zero.
Thus, in this $V=\infty$ limit, we find an ideal
insulator at any temperature independently of the integrability
of the system.

Now, allowing crossing of solitons and antisolitons leads
to a $D\neq 0$ in a finite size system. However our numerical results above
suggest that $D$ scales again to zero exponentially with the system size.
To estimate the dependence of the energies in the flux $\phi$ due to
$s\bar s$ crossing we can argue in terms of ordinary perturbation theory.
As above, a $\phi$ dependence will appear if by applying the
Hamiltonian on a $s\bar s$ state we can bring it to an equivalent one,
accumulating in the process a nonzero phase factor.
This process, that is now possible, necessitates the successive creation
and annihilation of $N_s$
$s\bar s$ pairs and will appear in the $\sim N_s$ order in perturbation
expansion. As it involves $N_s$ intermediate states with energy difference $V$
it is an exponentially small contribution.

Further, taking the point of view that the large $V/t$ limit is the fixed
point of the insulating behavior, we can argue against
the scenario of an insulator at zero temperature
developing to an ideal conductor at finite temperatures.

The next point to discuss is the possibility
of an integrable insulating system developing into a normal
conductor (resistor) at finite temperatures.
Unfortunately this point can only be settled
after clarifying the exact connection between
the static conductance, $\sigma_0$, and integrability.
However, consistently with our conjecture, we have found from independent
spectral analysis that whenever the insulating system is integrable,
the level spacing distribution is Poisson and $\sigma_0$ seems to vanish,
while when the system is nonintegrable,
the statistics is GOE and $\sigma_0\neq 0$.

Finally, in order to visualize the above picture of the insulating state,
we present in Fig. ~5 calculations in the large $V/t$ limit.
For $V=8t$ the system is characterized by a much larger charge gap
$\Delta_0 \sim 6t$.
Indeed we see a region of finite conductance in the frequency region
$0<\omega< 4t$ corresponding to excitations within the first,
one soliton-antisoliton pair,  "Hubbard" band. The weight in this region is
increasing exponentially with temperature, a sign of activated transport.
It is followed by a vanishing conductance up to $\omega \sim 6t$,
when transitions from the ground state to the first Hubbard band start.

{\it Comments:}
Our conclusions should hold for other integrable systems as well.
Since the anisotropic (and isotropic) spin-1/2 Heisenberg model is
equivalent to the integrable ($W=0$) model (\ref{eq5}), analogous conclusions
should apply for the spin stiffness and spin diffusion at $T>0$. Furthermore,
we have numerical evidence (to be presented elsewhere) that also
the integrable 1D Hubbard model, exhibits the same features found for the
prototype model (\ref{eq5}). Namely, out of half filling, the system seems to
be an ideal conductor, while at half-filling results are consistent with
an ideal insulator with $\sigma_0(T>0)=0$, for any strength of the repulsive
interaction.

We should stress that the above results are only
indicative of a relation between integrability and finite temperature
conductance and our arguments are far from rigorous.
Further analytical and numerical studies are necessary to prove
the validity of our conclusions.
However, taking into account present limitations on the size
of the systems that can be numerically studied and the absence of
analytical work, we think that the results
presented here are {\it qualitatively} clear enough to warrant further work
on this novel idea. Furthermore, an experimental effort is necessary to
observe this unusual conductance enhancement. Finally the stability of this
effect to deviations from integrability should be studied, similarly to the
problem of stability of classical soliton systems.

\acknowledgments
We would like to thank J. Jakli\v c for the help with the
Lanczos method and H. Castella for useful discussions. This work was
supported by the Swiss National Fond Grants No. 20-39528.93,
University of Fribourg and by the Ministry of Science and Technology
of Slovenia.

\begin{figure}

\caption{Integrated conductance $I(\omega)$ for $V=1.5t, W=0, T=2t$,
for $L = 12-16$ (exact diagonalization - full fines) and $L =20,24$
(Lanczos method - dotted lines).  Inset shows normalized charge
stiffness $D^*$ vs. $1/L$: exact diagonalization (disks), $T>0$ Lanczos
method (circles), analytical result (square) and $3^{rd}$ order
polynomial extrapolation from $L=8,12,16$ (dotted line).}
\label{one}

\caption{Integrated conductance $I(\omega)$ for $V=1.5t, W=t, T=2t$,
for $L=12 - 20$.  In the inset $ln D^*$ vs. $L$ is plotted. Notation is
as in Fig.~1.}
\label{two}

\caption{$I(\omega)$ for $L=16, V=1.5t, T=2t$, and $W/t = -0.5, -0.3, 0.,
0.3, 0.5$.}
\label{three}

\caption{$I(\omega)$ for $V=4t, W=0, T=4t$, for $L=12,16$
(full lines with increasing
line thickness-exact diagonalization) and $L=20$ (Lanczos method-dotted
line). Inset: $I(\omega)$ for $L=16$, $T=4t$.}
\label{four}

\caption{$I(\omega)$ for $V=8t, W=0, T=4t$, for $L=12,16$
with the notation as in Fig.~4.}
\label{five}

\end{figure}

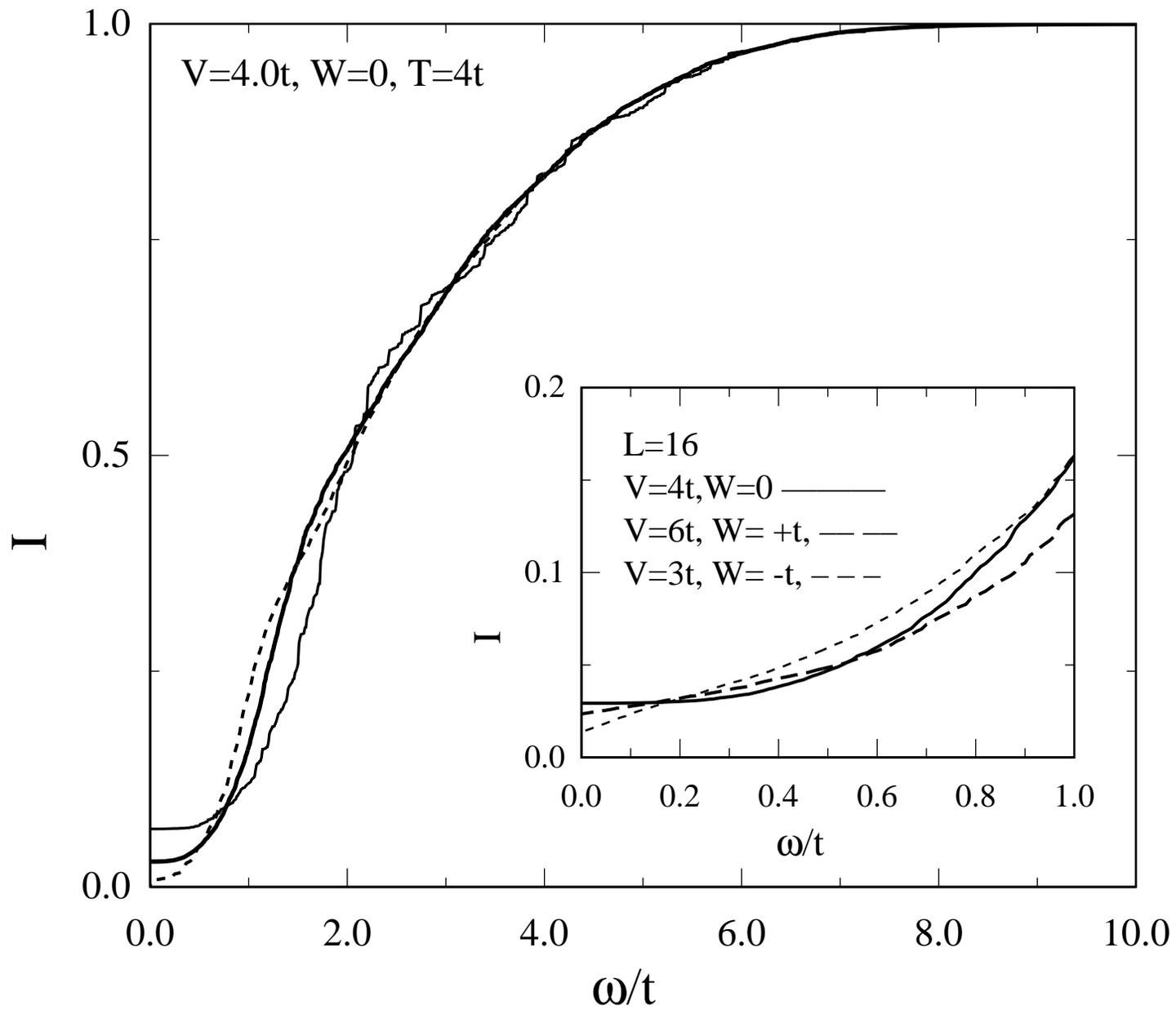
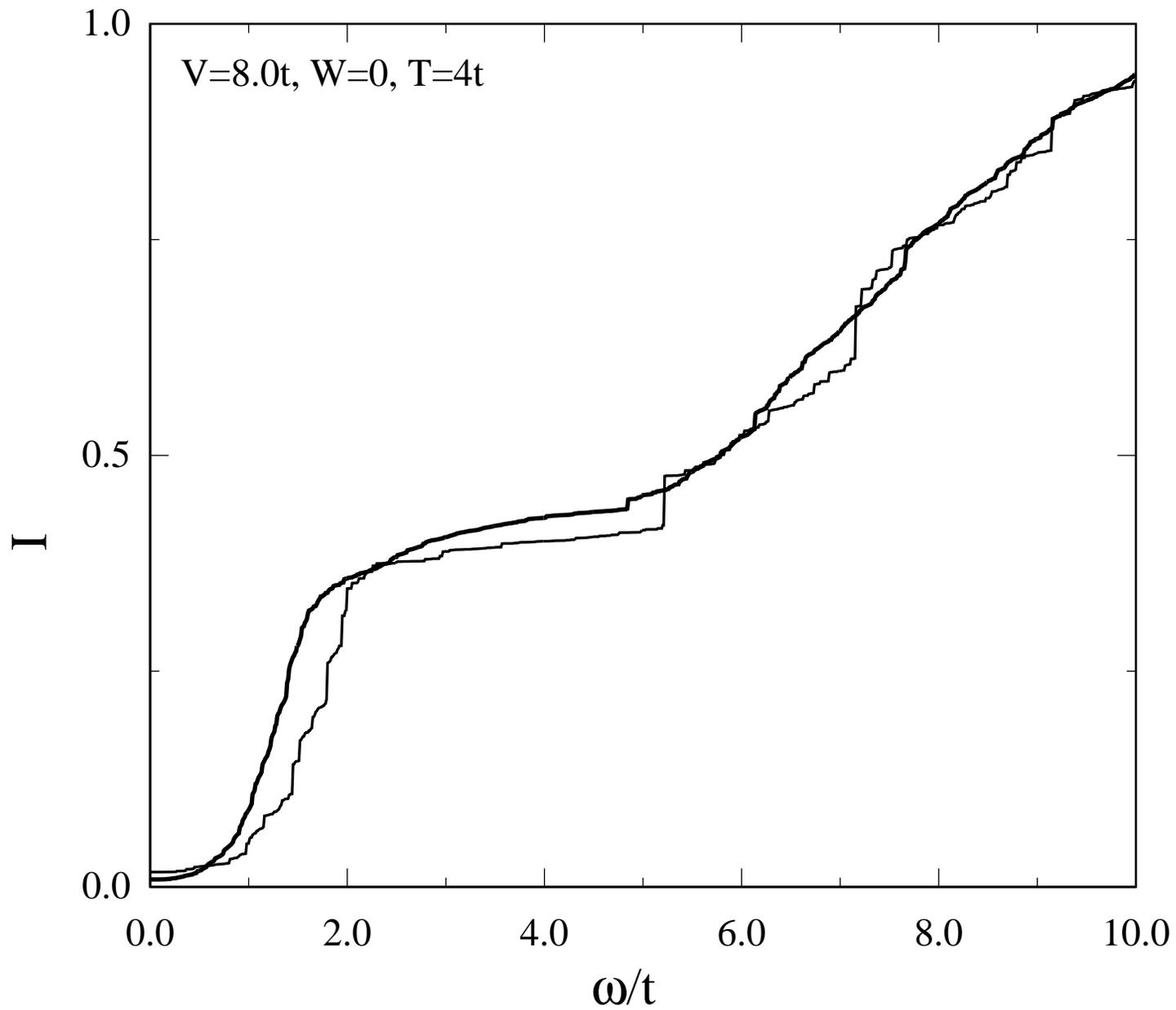
\end{document}